\documentclass[superscriptaddress,onecolumn,showpacs,
amssymb,amsmath,nobibnotes,aps,prd,
showkeys,
nofootinbib]{revtex4-2}
\pdfoutput=1
\usepackage{graphicx,subfigure,bm,color,psfrag,hyperref}
\usepackage{amsfonts}
\usepackage{lipsum}
\usepackage{mathtools}
\usepackage{verbatim}
\usepackage{color}
\usepackage{enumitem}

\begin{document}
\title{Alternative approaches to the description of quantum dynamics in multi-well potentials}
\author{V. P. Berezovoj}
\email{berezovoj@kipt.kharkov.ua}
\author{Yu. L. Bolotin}
\email{ybolotin@gmail.com}
\author{V. A. Cherkaskiy}
\email{vcherkaskiy@gmail.com}
\author{M. I.  Konchantnyi}
\email{konchatnij@kipt.kharkov.ua}
\affiliation{NSC ''Kharkov Institute of Physics and Technology'', Akademicheskaya Str. 1,	Kharkiv, 61108,	Ukraine}
\affiliation{V.N.Karazin Kharkiv National University, 61022 Kharkov, Ukraine}
\begin{abstract}
We consider three different approaches to analyze the quantum mechanical problems in multi-well potentials:
\begin{itemize}
    \item the standard matrix diagonalization technique in the basis sets of harmonic oscillator eigenfunctions or plain waves;
    \item the spectral method, which allows to reconstruct the spectrum and stationary functions based on the time-dependent solution of the Schr\"odinger equation;
    \item approximations with exact solutions obtained by the supersymmetric quantum mechanics technique.
\end{itemize}
The latter approach proves to be the most promising as it gives a unique possibility to include the specific multi-well features of the problem directly in the calculation procedure.
\end{abstract}
\maketitle
\tableofcontents
\section{Introduction}
The basic subject of the current investigation are Hamiltonian systems with potential energy surface which has several local minima, i.e. multi-well potentials. Such systems represent a realistic model, describing the dynamics of transition between different equilibrium states, including such important cases as chemical and nuclear reactions, nuclear fission, phase transitions, string landscape and many others.

The description of these processes requires the use of adequate mathematical methods in both classical and quantum mechanics. The traditional method for solving the quantum-mechanical problem (finding the spectrum and wave functions) is the diagonalization procedure. The effectiveness of the diagonalization procedure (the dimension of the Hilbert space used) depends to a large extent on the choice of basis. The choice of the latter is complicated by an extremely meager set of exactly solvable problems for multi-well potentials.

Methods that represent an alternative to the diagonalization procedure can be divided into two groups. The first includes methods that provide a direct alternative to the diagonalization procedure. An example of such a method is the so-called spectral method. In this method, the diagonalization procedure is replaced by a numerical solution of the non-stationary Schr\"odinger equation. The second group includes methods that increase the efficiency of the diagonalization procedure. These methods offer recipes for constructing an ``optimal'' basis for finding the spectrum and wave functions of a specific Hamiltonian. The basis is optimal on the set of exactly solvable problems with multi-well potentials. An extension of this set can lead to a change in the optimal basis.

Among the discussed methods, the method based on extended supersymmetric quantum mechanics (SQM) occupies a special place. Depending on how it is used, it can be included in both the first and the second group. On the one hand, it provides a mechanism for constructing exactly solvable problems for one-dimensional multi-well potentials, and on the other, these solutions can be used as an effective basis for solving problems that do not allow an exact solution.

In this paper, we consider the spectral method and the method based on the extended SCM, and give examples of their implementation.

\section{The Matrix Diagonalization Technique}
Let us consider the Schr\"odinger equation for a system with discrete energy spectrum

\begin{equation}
\label{eq5_1}
H\Psi_n=E_n\Psi_n,\ n\in \mathbb{N}
\end{equation}
and let there be full orthonormal basis of functions

\begin{equation*}
\varphi_k,\ k\in\mathbb{K},\ \langle\varphi_{k`}|\varphi_{k``}\rangle=\delta_{k`k``}
\end{equation*}
where $\mathbb{N}$ and $\mathbb{K}$ are countable sets.

The basis functions $\varphi_k$ are solutions of another Schr\"odinger equation

\begin{equation*}
h\varphi_k=e_k\varphi_k,\ k\in\mathbb{K}.
\end{equation*}
They are given analytically or are obtained numerically in an independent way.

Obviously, there exists a decomposition

\begin{equation*}
\Psi_n=\sum\limits_{k\in\mathbb{K}}a_k^{(n)}\varphi_k,\ a_k^{(n)}=\langle\varphi_{k}|\Psi_n\rangle.
\end{equation*}
The solution of the Schr\"odinger equation (\ref{eq5_1}) by the matrix diagonalization technique implies the following:

\begin{enumerate}
	\item the set $\mathbb{K}$ is presented as a direct sum of the subsets
	
	\[\mathbb{K}=\bar{\mathbb{K}}\oplus\mathbb{K}'\]
	such as $\bar{\mathbb{K}}$ is finite and $\mathbb{K}$ is a countable set.
	
	\item the original Hamiltonian of the problem (\ref{eq5_1}) is presented in the form
	
	\[H=\bar{H}+H',\]
	where by definition
	
	\[\langle\varphi_{k'}|\bar{H}|\varphi_{k''}\rangle=\langle\varphi_{k'}|H|\varphi_{k''}\rangle,\ k', k''\in\bar{\mathbb{K}}\]
	and all other matrix elements of $\bar{H}$ are zeros.
	\item the eigenvalue problem 
	\[\bar{H}\bar{\Psi}_n=\bar{E}_n\bar{\Psi}_n,\ n\in \mathbb{K}\]
	is solved numerically, where
	\[\bar{\Psi}_n=\sum\limits_{k\in\bar{\mathbb{K}}}\bar{a}_k^{(n)}\varphi_{k},\ \bar{a}_k^{(n)}=\langle\varphi_{k}|\Psi_{n}\rangle,\ \langle\bar{\Psi_{n'}}|\bar{\Psi_{n''}}\rangle=\delta_{n'n''}\]
\end{enumerate}
Harmonic oscillator and infinitely deep potential well in fact exhaust the set of exactly solvable one-dimensional quantum systems whose eigenfunctions can be used as a basis for matrix diagonalization. There are many more possibilities in the models with dimensionality of more then one. Further, for simplicity we consider only two-dimensional systems, but the results can be trivially generalized for higher dimensions.

The simplest type of two-dimensional basis can be constructed from products of eigenfunctions of exactly solvable one-dimensional problems, for example, from plane waves
\[\varphi_{k_x,k_y}(x,y;a_x,a_y)=\frac1{\sqrt{a_xa_y}} \sin\frac{\pi k_x}{2a_x}(x+a_x) \sin\frac{\pi k_y}{2a_y}(x+a_y),\]
harmonic oscillator eigenfunctions
\[\varphi_{k_x,k_y}(x,y;\omega_x,\omega_y)= \frac{(\omega_x\omega_y)^{1/4}}{\sqrt{\pi\hbar}} \frac{H_{k_x}\left(\sqrt{\frac{\omega_x}\hbar}x\right) H_{k_y}\left(\sqrt{\frac{\omega_y}\hbar}y\right)}{\sqrt{2^{k_x+k_y}k_x!k_y!}} \exp\left(-\frac{\omega_x x^2+\omega_y y^2}{2\hbar}\right),\]
or as a combination of both of them
\[\varphi_{k_x,k_y}(x,y;a,\omega)=\frac1{\sqrt{n}} \left(\frac{\omega}{\pi\hbar}\right)^{1/4} \sin\frac{\pi k_x}{2a}(x+a) \frac{H_{k_y}\left(\sqrt{\frac{\omega}\hbar}y\right)}{\sqrt{2^{k_y}k_y!}}, \exp\left(-\frac{\omega y^2}{2\hbar}\right).\]
\section{The Spectral Method}
The spectral method (SM) for the solution of the Schr\"odinger equation was proposed in the paper [44] in application to 1D and 2D potential systems, but it can be easily generalized for the Schr\"odinger equations of arbitrary dimensions:

\[\left[-\frac{\hbar^2}{2}\sum\limits_{i=1}^{D}\partial_i^2 + U(x_1,\dots,x_D)\right] \psi_n (x_1,\dots,x_D) = E_n \psi_n (x_1,\dots,x_D),\]
where D is the dimensionality of the system's configuration space. Let us assume that the potential $U(x_1,\dots,x_D)$ allows only finite motion for all energies, therefore our task is to find discrete energy spectrum $E_n$ and stationary wave functions $\psi_n(x_1,\dots,x_D)$.

Let us consider time-dependent solution $\psi (x_1,\dots,x_D;t)$ for the corresponding non-stationary Schr\"odinger equation

\[\left[-\frac{\hbar^2}{2}\sum\limits_{i=1}^{D}\partial_i^2 + U(x_1,\dots,x_D)\right] \psi(x_1,\dots,x_D;t) = i\hbar\partial_t \psi(x_1,\dots,x_D;t)\]
with some in principle arbitrary initial condition

\[\psi_0
(x_1,\dots,x_D)=\psi_n (x_1,\dots,x_D;t=0)\]
Applying the decomposition

\[\psi_0 (x_1,\dots,x_D)=\sum\limits_{n=1}^{\infty}a_n \psi_n (x_1,\dots,x_D),\]
we obtain
\[\psi (x_1,\dots,x_D;t)=\sum\limits_{n=1}^{\infty}a_n \psi_n (x_1,\dots,x_D)e^{-i\frac{E_n t}{\hbar}}.\]
\section{Comparative analysis of matrix diagonalization and spectral methods}
The method of Hamiltonian diagonalization is the most traditional way for numerical solution of the Schr\"odinger equation. The spectral method for solution of the same problem represents in its turn a newer one and for many reasons a more preferable approach. One of the most fundamental disadvantages of the matrix diagonalization technique is the rather poor choice of exactly solvable models whose eigenfunctions can be taken as a basis for subsequent diagonalization of the Hamiltonian under consideration. As a rule, properties of the Hamiltonian pose very rigid limitations on the auxiliary basis parameters, therefore in most cases of matrix diagonalization implementation, only one free parameter remains---it is the auxiliary basis dimension $N$. Further it is necessary only to determine the minimal dimensionality $N$ sufficient for the achievement of desired resulting accuracy. Such simplicity of the matrix diagonalization method results in its insufficient  flexibility: in practice the application of matrix diagonalization is justified  only for those potentials that can be approximated at least locally by some exactly solvable model. But such a limitation cannot be satisfied for many important problems, especially in the  potentials with many local minima.

The spectral method uses a natural basis of free particle wave
functions---such a basis is equally good, or better to say equally
bad, for potentials of any form. Such fundamental indifference of
the spectral method to shapes of potential energy surface is the
main reason of its universality. Compared to matrix diagonalization,
the spectral method has much more flexibility---the researcher is
free to choose both the length and step of the computation grid in
time ($T$ and $\Delta t$) as well as in coordinate space ($L_i$ and
$\Delta x_i$). In the same time choice of the nodes number $N$ is
limited by the computational efficiency requirement: the
applicability condition for the fast Fourier transform algorithm
--- the main basis of the spectral method efficiency --- assumes that
all $N_i$ do not contain large simple factors; ideally all of them
should be integer powers of two $(N_i=2^{k_i})$. And, last but not
least, the main freedom lies in the choice of the initial state for
the spectral method computations. As it is very difficult to give
any general recommendation on that point, the spectral method
computations have become a real art rather than plain technique,
requiring great experience and constant practice. Because the
spectral method is not standardized up to the present time, the fast
Fourier transform represents the only one ready-to-use ingredient
for its realization, available in many well-known software
libraries. Other stages of computations require rather extended
although principally simple software development. On the other hand,
the spectral method algorithm itself can be easily generalized for
problems of any dimensions, which is not the case with the matrix
diagonalization technique---reasonable construction of finite
multi-dimensional basis from one-dimensional eigenstates always
represents a non-trivial task because of the basis vectors ordering
problem.

A quantitative measure of the numerical method efficiency is the
growth of computational expense --- CPU time and RAM usage --- with
increase in the results both quantity and quality. It is useful to
compare the matrix diagonalization and the spectral method
efficiencies in calculation of $n$ energy levels of a quantum system
with fixed relative error $\varepsilon$. Taking into account the
fact that for studies of statistical properties the calculated
spectrum is inevitably unfolded, it is reasonable to define the
accuracy as the maximum ratio of absolute error of the computed
energy levels to mean level spacing
\[\varepsilon=\frac{\delta_E}{\Delta_E}=\rho(E)\delta_E.\]

For most potential systems the level density $\rho(E)$ grows quite
fast with the energy:
\[\rho(E)\approx E^{\lambda E-1},\]
where $D$ is the system dimensionality and $\lambda$ is close to
unity (it exactly equals unity for a harmonic oscillator). Therefore
the condition to achieve the desired accuracy will be the most
critical for levels with maximum energy, while the lowest levels
will be obtained with higher accuracy than needed --- this
inconvenient feature is due to the very nature of smooth potential
systems and is equally shared by both methods under consideration.

In the matrix diagonalization method the absolute computational
error for sufficiently low energies does not exceed the round-off
errors:
\[\delta_{E_k}\approx\delta_0 E_k N^2,\ k<\eta N, 0<\eta<1\]
where $N$ is the basis dimensionality, $\eta$ represents the
problem-dependent rate of correctly calculable states and $\delta_0$
is the machine round-off error ($\delta_0\sim10^{-15}$ for standard
double accuracy numbers). For such levels the relative error
$\varepsilon\approx E^{\lambda D}\delta_0 N^2$ appears to be
negligibly small and for required basis dimensions we get the
condition:
\[N_{MD}=\frac n \eta.\]

In the general case of matrix diagonalization the computation time
scales as $T_{MD}\approx N^3=n^3/\eta^3$ and memory usage is
$M_{MD}\approx N^2=n^2/\eta^2$. However in many important particular
cases, for example for polynomial potentials, the matrix
diagonalization implementation involves band matrices, and in such
cases $T_{BMD}\approx N^2$ and $M_{BMD}\approx N$.

In the spectral method the computational accuracy is determined by
the size of time computational grid: $\delta_E\approx1/T$, while the
time step determines the spectral bandwidth where the levels can be
determined: $\Delta t\approx 1/E_{max}$. Therefore the maximum
relative error in the spectral method scales as $\varepsilon\approx
E^{\lambda D-1}/T$, and the required time step number $N_T=T/\Delta
t\approx n/\varepsilon$. Because the spectral method actually uses
the plane waves decomposition for the computed wave functions, the
required nodes number for the computational grid for each of the
dimensions, equal to the effective basis vector number on the same
degree of freedom, is determined in the same way as for the
corresponding basis in the matrix diagonalization method:
\[N_i\approx \left(\frac{n}{\eta_{PW}}\right)^{\frac1D},\]
where $\eta_{PW}$ is the rate of correctly calculable levels on the
plane waves basis. Therefore the computation time scales as
\[T_{SM}\approx N_T N_i^D\ln N_i\approx\frac{n^2}{\varepsilon\eta_{PW}}\ln\frac{n}{\eta_{PW}}\]
and memory usage scales as
\[M_{SM}\approx N_i^D\approx\frac{n}{\eta_{PW}}\]
As a result the spectral method is generally more efficient for both
CPU time and RAM usage criteria:
\[\frac{M_{SM}}{M_{MD}}\approx\frac{\eta^2}{n\eta_{PW}},\ \frac{T_{SM}}{T_{MD}}
\approx\frac{\eta^3}{n\varepsilon\eta_{PW}}\ln\frac{\eta}{\eta_{PW}}.\]

The last but not least advantage of the spectral method lies in the
fact that numerical simulation of wave packets temporal dynamics is
included into the method algorithm as an auxiliary procedure. In
some important applications this simulation represents the ultimate
goal of research, while the determination of energy levels and
stationary wave functions is not required. In such cases an
important advantage of the spectral method is the possibility of
achieving the goal by the shortest way, while it would require much
more computational efforts to reproduce the same results by the
matrix diagonalization technique.
\section{Spectral method efficiency example: signatures of quantum chaos in the wave function structure}
In this section, we will demonstrate the effectiveness of the spectral method for studying the quantum manifestations of classical stochasticity (QMCS) in the wave functions structure, which can be realized in potentials with two and  more local minima. Efficiency of the approach is demonstrated for two potentials: surface quadrupole oscillations (QO) and lower umbillic catastrophe (UC) $D_5$.

After almost a hundred years of development, quantum mechanics became a universal picture of the world. On any observable scales of energy we could not find any violations of quantum mechanics. But this does not mean that from time to time quantum mechanics does not confront another challenge. The problem that arose at the face of quantum mechanics in the second part of the last century is called quantum chaos. The essence of the problem is the fact that, on the one hand, the energy spectrum of any quantum system with finite motion is discrete and thus its evolution is quasi-periodic, but, on the other hand, the correspondence principle demands transition to classical mechanics which demonstrates not only regular solutions but chaotic too.  In other words, the problem consists in the fact that the discrete nature of the spectrum never implies chaos, or more exactly any resemblance to chaos in the sense of the ergodicity theory, in any quantum system with finite motion. Meanwhile the correspondence principle in the semiclassical limit requires the presence of chaos, connected with the nature of motion in the classical case.

The problem of quantum chaos has not yet received a final solution. However, an alternative approach to the problem of quantum chaos is also possible. Not waiting for the complete solution of the problem (or rather for its correct formulation) we can study its limited variant: investigation of special features of quantum systems whose classical analogues are chaotic, i.e. to search quantum manifestations of classical stochasticity. One of the fragments of this problem will be considered below.

Energy spectra and eigenfunctions of classically nonintegrable systems represent the main object of search for QMCS [1, 2, 3]. It should be pointed out that in analysis of QMCS in the energy spectra the principal role was given to statistical characteristics, i.e. quantum chaos was treated as property of a group of states. In contrast, the choice of a stationary wave function as a basic object of investigation, relates quantum chaos to an individual state. Usual procedure of search for QMCS in wave function implies investigation of distinction in its structure below and above the classical energy of transition to chaos (or other parameters of regularity-chaos transition). Such procedure meets difficulties connected with necessity to separate QMCS from modifications of wave functions structure due to trivial changes in its quantum numbers. Up to present time correlations between peculiarities of the classical motion and structure of wave functions were studied mostly for billiard-type systems [4, 5, 6]. For Hamiltonian systems with non-zero potential energy QMCS were studied either for model wave functions [7] or for potential energy surfaces (PES) with simple geometry [8]. Till now there is practically no information on wave functions structure for generic Hamiltonian systems, including multi-well potentials. Such systems allow existence of the mixed state (MS): different (regular or chaotic) classical regimes coexist in different local minima at fixed energy [9, 10]. Such systems represent optimal object for investigation of QMCS in wave functions structure. Wave functions of MS allow to find QMCS in comparison not different eigenfunctions, but different parts of the same wave function, situated in different regions of configuration space (corresponding to different local minima of the potential). Let us demonstrate this possibility for MS, generated by the deformation potential of surface QO of atomic nuclei [11] and lower UC D5 [12]. It can be shown [13], that using only the transformation properties of the interaction, the QO potential takes the form

\begin{equation}
\label{Uaa} U_{QO}(a_0,a_2)=\sum_{m,n}C_{mn}(a_0^2+2a_2^2)^m a_0^n
(6a_2^2-a_0^2)^n,
\end{equation}
where $a_0$ and $a_2$ are internal coordinates of nuclear surface
undergoing the QO
\begin{equation}
\label{Rtf} R(\theta,\varphi)=R_0\{1+a_0
Y_{2,0}(\theta,\varphi)+a_2
[Y_{2,2}(\theta,\varphi)+Y_{2,-2}(\theta,\varphi)]\}
\end{equation}

\begin{figure}
	\includegraphics[width=\textwidth]{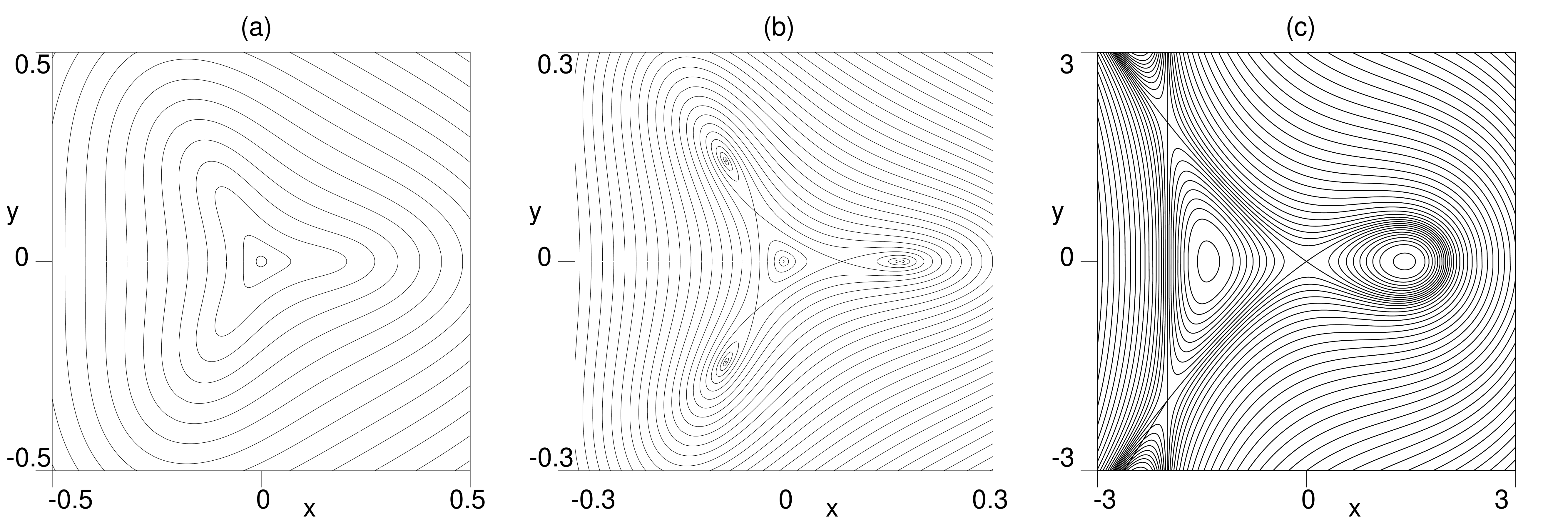}
	\caption{The level lines of the QO potential
		(\ref{Uxydl}) for $W=13$ (a), $W=18$ (b), and for the UC $D_5$
		(\ref{d5}) with $a = 2$ (c). \label{w13w18d5}}
\end{figure}

Restricting ourselves with the terms of fourth order in the deformation and assuming equality of masses for the two independent directions, we get the following  $C_{3v}$-symmetric Hamiltonian 

\begin{subequations}
	\begin{equation}
	\label{Hxydl} H = \frac{p_x^2+p_y^2}{2m}+U_{QO}(x,y;W)
	\end{equation}
	\begin{equation}
	\label{Uxydl} U_{QO}(x,y;W) = \frac{1}{2W}(x^2+y^2)+xy^2-\frac 1 3
	x^3+\left(x^2+y^2\right)^2
	\end{equation}
\end{subequations}
where

\[x = a_0, y = \sqrt{2}a_2, W=\frac{b^2}{ac},\]
\[a = 2C_{10}, b = 3C_{01}, c = C_{20}.\]
Hamiltonian (\ref{Hxydl}) and corresponding equations of motion depend only on $W$, which is the unique dimensionless parameter, that can be constructed from $a$, $b$ and $c$, and it completely determines the potential energy surface (PES) [Fig.\ref{w13w18d5}(a),(b)]. Region $0<W\le16$ includes potentials with only one critical point---minimum in the origin [Fig.\ref{w13w18d5}(a)], corresponding to spherically symmetric equilibrium shape of the nucleus (or liquid charged drop). For $W>16$ the PES has seven critical points: four minima (one central and three peripheral) and three saddles, separating the peripheral minima from the central one [Fig.\ref{w13w18d5}(b)]. Below  we consider in details the case $W=18$, when the potential (\ref{Uxydl}) has four minima with the same value $E_{min}=0$ and the saddle energies $E_S=1/20736$. It was shown [9], that the critical energy of transition to chaos $E_{cr}$ has different values for different minima: $E_{cr}=E_S/2$ for the central minimum and $E_{cr}=E_S$ for the peripheral ones. It means, that for $E_S/2<E<E_S$ regular and chaotic trajectories coexist and are separated not in phase, but in configuration space, resulting in the phenomenon of MS.

MS is a common case for multi-well potentials. According to catastrophe theory, a wide class of multi-well 2D polynomial potentials can be generated by germs of lower UC of types $D_5,D^\pm_6,D_7$ from the Thom catastrophes list, affected by certain perturbation [12]. We consider a lower UC $D_5$, described by the germ $x^4/4+y^2x$ with the perturbation $bx^2-ay^2$. This potential has only two local minima and three saddles [Fig.\ref{w13w18d5}(c)], and therefore it is the simplest potential, where MS is observed. Under the Maxwell condition $b=a^2/4$ the energies of all the saddles are the same, and energies of all the local minima too. We will consider the case $a=2,b=1$
\begin{equation} \label{d5} U_{D_5}(x,y) =
\frac{x^4}{4}+xy^2+2y^2-x^2
\end{equation}
when all $E_{min}=-1$ and all $E_S=0$, and MS is observed in the energy region $-1/2<E_{MS}<0$.

As we noted above, the calculation of quasiclassical part of the spectrum for systems with multi-well PES requires appropriate numerical methods. In this case the attractive alternative to the matrix diagonalization may become the spectral method.

\begin{figure}
	\includegraphics[width=\textwidth]{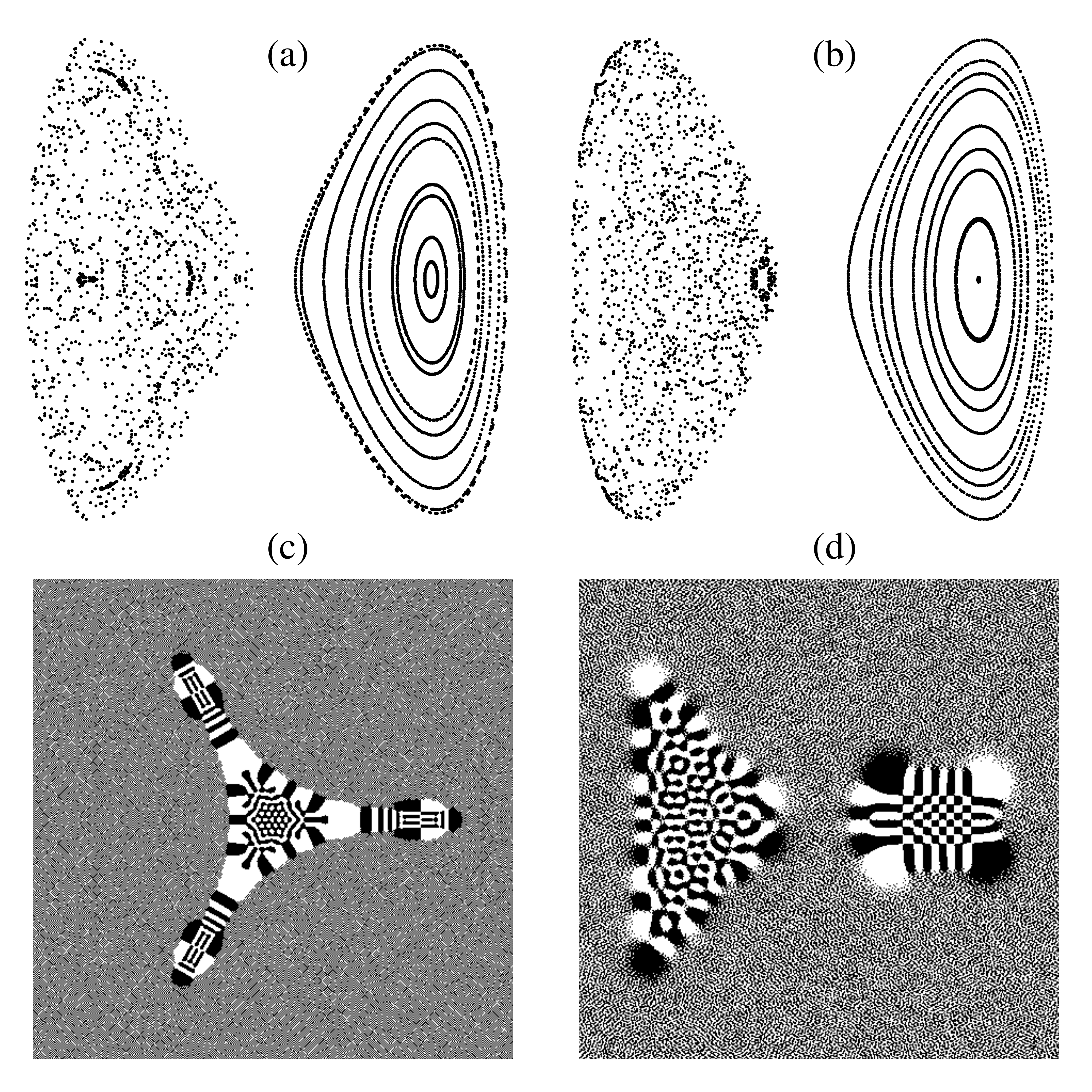}
	\caption{The MS in the QO potential (a),(c) and in the
		$D_5$ UC (b),(d): (a),(b) -- Poincar\'e surfaces of section,
		(c),(d) -- nodal domains of the eigenfunctions. \label{w18d5}}
\end{figure}

We now turn to the analysis of the results obtained using the spectral method for potentials QO and $D_5$. Existence of the MS, at $W>16$ for the QO potential  [Fig.\ref{w18d5}(a)] or in the $D_5$ potential [Fig.\ref{w18d5}(b)], opens a new possibility for investigation of QMCS. Comparing the structure of the eigenfunction in central and peripheral minima of the QO potential [Fig.\ref{w18d5}(c)], or in the left and the right minima of the UC $D_5$ potential [Fig.\ref{w18d5}(d)], it is evident that the nodal structures of the regular part and the chaotic part of the eigenfunction are clearly different:

\begin{enumerate}[label=(\roman*)]
	\item within the classically allowed region the nodal domains of the regular part of the wave function form a well recognizable checkerboard-like pattern [1]; nothing similar can be observed for the chaotic part;
	\item the nodal lines of the regular part exhibit crossings or very tiny quasicrossings; in the chaotic part the nodal lines quasicrossings have significantly larger avoidance ranges;
	\item while crossing the classical turning line $U(x,y)=E_n$, the nodal lines structure of the regular part immediately switches to the straight nodal lines, going to inﬁnity, which makes the turning point line itself easily locatable in the nodal domains structure; in the chaotic part an intermediate region exists around the turning line, where some of the nodal lines pinch-oﬀ, making transition to the classically forbidden region more graduate and not so manifesting in the nodal structure.
\end{enumerate}
\section{Multi-well potentials and extended supersymmetric quantum mechanics}
Hamiltonians with multi-well potentials effectively model important features of both classical and quantum dynamics. The spectrum and wave functions of quantum systems of this type can be (in principle) obtained by diagonalization of the corresponding Hamiltonians. An essential criterion for the effectiveness of the diagonalization procedure is an adequate choice of basis. The presence of exactly solvable [8] or partially solvable [9] quantum-mechanical models with multi-well potentials makes it possible to achieve significant progress in constructing bases compatible with available computational capabilities. An effective generator of new exactly solvable quantum models can be SUSY QM.
\subsection{N = 4 SQM and multi-well potentials}
Historically supersymmetric quantum mechanics (SQM) appeared as a very simple and simultaneously very efficient laboratory for the investigation of supersymmetry and its consequences in particle physics [10]. In turn, the presence of some number of supercharges commuting with the Hamiltonian determines the structure of the SQM Hamiltonian in terms of usual coordinates and momenta and additional anticommuting variables, which can be represented by matrices. Due to this matrix structure, the diagonalized Hamiltonian of SQM consists of a number of normal quantum mechanics Hamiltonians that are mathematically related. In the case of $N = 2$ SQM these relations comprise the well known Darboux transformations [11]. If the number of supercharges grows, such extended supersymmetry leads to additional interesting consequences and in the case of $N = 4$ (or $N = 2$ if we consider complex supercharges) SQM is very closely related to the inverse scattering problem [12,13].

Extended $N = 4$ SQM [12,13] is equivalent to second-order polynomial supersymmetric quantum mechanics (SQM) (reducible case) [14,15] and assumes the existence of complex operators of supersymmetries $Q_1(\bar{Q_1})$ and $Q_2(\bar{Q_2})$, through which the Hamiltonians $H_{\sigma_1}^{\sigma_2}$ can be expressed. The Hamiltonian of $N = 4$ SQM has a form ($\hbar=m=1$)
\begin{equation}
\begin{array}{c}
\displaystyle H_{\sigma_1}^{\sigma_2}=\frac{1}{2}(p^2+V_2^2(x)+\sigma_3^{(1)}V'_2(x))\equiv \frac{1}{2}(p^2+V_1^2(x)+\sigma_3^{(2)}V'_1(x)), \\ 
\displaystyle V_i(x)=W'(x)+\frac{1}{2}\sigma_3^{(i)}\frac{{W''(x)}}{{W'(x)}}
\end{array}
\label{susy-article-eqs:11}
\end{equation}
where $W(x)$ is a superpotential and $\sigma_3^{(i)}$ -- matrices, which commute with each other and have eigenvalues $\pm 1$, $\sigma_3^{(1)}= \sigma_3 \otimes 1, ~\sigma_3^{(2)}= 1 \otimes \sigma_3$. 

Supercharges $Q_i$ of extended supersymmetric quantum mechanics form an algebra: 
\begin{equation}
\begin{array}{c}
\left\{{Q_i, \bar Q_k}\right\}= 2\delta_{ik}H, ~\left\{{Q_i, Q_k}\right\}= \left\{{\bar Q_i, \bar Q_k}\right\}= 0, ~i, k = 1, 2\\
Q_i = \sigma_-^{(i)}(p+iV_{i+1}(x)), ~\bar Q_i = \sigma_+^{(i)}(p - iV_{i+1}(x))\\ \end{array}
\label{susy-article-eqs:12}
\end{equation}
where $V_3(x)\equiv V_1(x), ~\sigma_{_\pm}^{(1)}= \sigma_\pm \otimes 1, ~\sigma_\pm^{(2)}= 1 \otimes \sigma_\pm$.

Hamiltonian and supercharges act on four-dimensional internal space and Hamiltonian is diagonal on vectors $\psi_{\sigma_1}^{\sigma_2}(x, E)$,  where $\sigma_1, ~\sigma_2$ -- eigenvalues of $\sigma_3^{(1)}, ~\sigma_3^{(2)}$. Supercharges $Q_i$($\bar Q_i$) act as lowering (raising) operators for indexes $\sigma_1, ~\sigma_2$. It is convenient to represent the structure of Hamiltonian and connection between wave functions in diagram form:
\begin{figure*}[h]
	\begin{center}
		\includegraphics{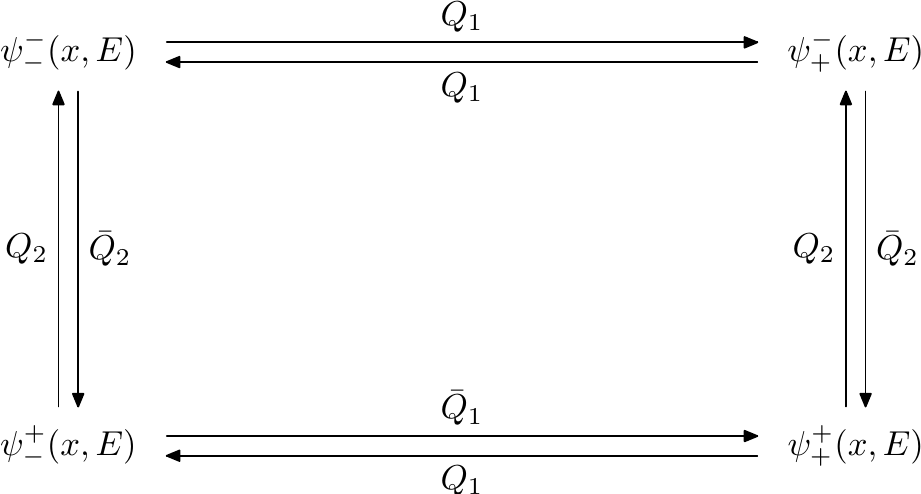}
	\end{center}
\end{figure*}
Obviously, since the operators $Q_i$ and $\bar Q_i$ are commutative with the Hamiltonian, all functions $\psi_{\sigma_1}^{\sigma_2}(x, E)$ are eigenfunctions of the Hamiltonian with the same eigenvalue $E$. The only exception is the case when the wave functions vanish under the action of supersymmetry generators.

Construction of isospectral Hamiltonians in the framework of $N = 4$ SQM is based on the fact that four Hamiltonians are combined into supermultiplet $H_{\sigma_1}^{\sigma_2}$. Nevertheless, it should be noted that due to symmetry $\sigma_1 \leftrightarrow \sigma_2 - H_{\sigma_1}^{-\sigma_2}\equiv H_{-\sigma_1}^{\sigma_2}$, only three of them are nontrivial. 

We will consider the procedure for constructing isospectral Hamiltonians with the addition of levels below the energy of the ground state of the initial Hamiltonian $\bar H_0$, which has only the states of the discrete spectrum. Particular attention will be paid to obtaining exactly solvable models with multi-well potentials.

Consider the auxiliary equation:
\begin{equation}
H_{\sigma_1}^{\sigma_2}\varphi(x)= \varepsilon \varphi(x)
\label{susy-article-eqs:13}
\end{equation}

As initial let's take one of the Hamiltonians 
\begin{equation}
\begin{array}{c}
H_{\sigma_1}^{\sigma_2}= \frac{1}{2}(p - i\sigma_1 V_2(x))(p+i\sigma_1 V_2(x))+\varepsilon \equiv \\
\equiv\frac{1}{2}(p - i\sigma_2 V_1(x))(p+i\sigma_2 V_1(x))+\varepsilon \\
\end{array}
\label{susy-article-eqs:14}
\end{equation}
where \[V_{\sigma_2}(x)=W'(x)+\frac12\sigma_2 W''(x)/W'(x)\] and $\varepsilon$ is the so-called factorization energy. In the future, the energy is counted from $\varepsilon$. Strictly speaking, the supersymmetry relations with supercharges of the form (\ref{susy-article-eqs:12}) and expressed through superpotential $W(x)$ are satisfied not by Hamiltonian $H$ itself, but by shifting by an amount of $\varepsilon$ Hamiltonian $H-\varepsilon$. However, due to commutativity of supercharges with constant $\varepsilon$, the relations between wave functions of operators $H$ and $H-\varepsilon$ are the same. If the operator $H_{\sigma_1}^{\sigma_2}$ is fixed, the form of $W(x)$ depends on the choice of factorization energy. Thereby the Hamiltonians also have nontrivial dependence on $\varepsilon$.

When $\varepsilon < E_0$ (where $E_0$ is the ground state energy of the initial Hamiltonian), the solutions of auxiliary equation (\ref{susy-article-eqs:13}) are  linearly independent functions $\varphi_i(x, \varepsilon), i = 1, 2$, which are non-negative and have the following asymptotic behavior: when $x \to -\infty\;\;\varphi_1(x)\to+\infty \;(\varphi_2(x)\to 0)$ and when $x \to+\infty \;\;\varphi_1(x)\to 0\;(\varphi_2(x)\to+\infty)$. Therefore. with appropriately chosen constants the general solution has the form $\varphi(x, \varepsilon, c)= N(\varphi_1(x, \varepsilon)+c\varphi_2(x, \varepsilon))$. ($N$ is common constant that will be used for normalization). Thus, the function \[\displaystyle\tilde \varphi(x, \varepsilon, c)= \frac{{N^{- 1}}}{{\varphi(x, \varepsilon, c)}}\] is finite and can be normalized at every concrete choice of $\varepsilon$ and $c>0$. Let us note that with concrete values of $\varepsilon$ and $c$, $\varphi(x, \varepsilon, c)$  can have local extreme points. In this case the natural choice of the initial Hamiltonian is $H_-^+$ or $H_+^-$ (which are identical due to symmetry of $H_{\sigma_1}^{\sigma_2}$ under $\sigma_1\leftrightarrow \sigma_2$). Then superpotential has the form [14]:
\begin{equation}
W(x, \varepsilon, \lambda)= - \frac{1}{2}\ln(1+\lambda \int\limits_{x_i}^x{dt}\tilde \varphi^2(t, \varepsilon, c))
\label{susy-article-eqs:15}
\end{equation}
with two new arbitrary parameters $\lambda, x_i$, but last of them is inessential, because it gives an additive contribution to $W(x)$. All Hamiltonians that form a supermultiplet have nontrivial dependence on these parameters.

Let us consider the connection between Hamiltonians from supermultiplet. For definiteness take $H_+^-$ as initial. We denote $\psi_+^-(x, E)$ the solution of the equation
\begin{equation}
H_+^- \;\psi_+^-(x, E)= E\psi_+^-(x, E)
\label{susy-article-eqs:16}
\end{equation}
Then using the first representation of Hamiltonian $H_-^-$ (\ref{susy-article-eqs:16}), we obtain the following relation for $H_-^-$, $\psi_-^-(x, E)$ and initial expressions: 
	\begin{equation}
\begin{array}{c}
\displaystyle H_-^- = H_+^-+\frac{{d^2}}{{dx^2}}\ln \tilde \varphi(x, \varepsilon, c),\\
\displaystyle  \psi_-^-(x, E_i)= \frac{1}{{\sqrt{2(E_i - \varepsilon)}}}\frac{{W\left\{{\psi_+^-(x, E_i), \varphi(x, \varepsilon, c)}\right\}}}{{\varphi(x, \varepsilon, c)}}\\
\displaystyle  \psi_-^-(x, E = 0)= \frac{{N^{- 1}}}{{\varphi(x, \varepsilon, c)}}= \tilde \varphi(x, \varepsilon, c)
\end{array}
\label{susy-article-eqs:17}
\end{equation}
Using the second representation $\displaystyle  H_{\sigma_1}^{\sigma_2}= \frac{1}{2}(p - i\sigma_2 V_{\sigma_1}(x))(p+i\sigma_2 V_{\sigma_1}(x))+\varepsilon$ and given equality $H_+^- \equiv H_-^+$, it is easy to bind $H_+^+$, $\psi_+^+(x, E)$ with original values:
\begin{equation}
\begin{array}{c}
\displaystyle H_+^+= H_+^- + \frac{{d^2}}{{dx^2}}\ln \left({\frac{{\tilde \varphi(x, \varepsilon, c)}}{{1+\lambda \int\limits_{x_i}^x{dt\;\tilde \varphi^2(t, \varepsilon, c})}}}\right)\;\quad,\\
\displaystyle \psi_+^+(x, E = 0)= \frac{{N_\lambda^{- 1}\tilde \varphi(x, \varepsilon, c)}}{{\;(1+\lambda \int\limits_{x_i}^x{dt\;\tilde \varphi^2(t, \varepsilon, c))}}}\\
\displaystyle \psi_+^+(x, E_i)= \frac{1}{{\sqrt{2(E_i - \varepsilon)}}}\left({\frac{d}{{dx}}+\frac{d}{{dx}}\ln \frac{{\tilde \varphi(x, \varepsilon, c)}}{{\;(1+\lambda \int\limits_{x_i}^x{dt\;\tilde \varphi^2(t, \varepsilon, c))}}}}\right)\psi_+^-(x, E_i)
\end{array}
\label{susy-article-eqs:18}
\end{equation}
It should be noted that normalization of the wave function $\psi_+^+(x, E =0)$ is possible for any $\tilde \varphi(x, \varepsilon, c)$. Using in the condition of that normalization relation
\begin{equation}
\label{susy-article-eqs:18n}
\displaystyle  \frac{{N_\lambda^{- 2}\tilde \varphi^2(x, \varepsilon, c)}}{{(1+\lambda N^{- 2}\int\limits_{- \infty}^x{dt\tilde \varphi^2(t, \varepsilon, c)})^2}}= - \frac{{N_\lambda^{- 2}}}{{\lambda N^{- 2}}}\frac{d}{{dx}}\frac{1}{{(1+\lambda N^{- 2}\int\limits_{- \infty}^x{dt\tilde \varphi^2(t, \varepsilon, c)})}}
\end{equation}
it is easy to get a connection between the normalization constants $N_\lambda^{-2}=(1+\lambda)N^{-2}$. The use of superpotential (\ref{susy-article-eqs:15}) with $\tilde \varphi(x, \varepsilon, c)$ corresponds to exact supersymmetry, which leads to the presence of zero modes in both $H_-^-$ and $H_+^+$.
\subsection{General properties of and multi-well potentials\label{5_2}}
We begin the consideration of the properties of the isospectral Hamiltonians obtained in the previous section without specifying the explicit form of the initial Hamiltonian. For this we use the formalism described above. Let $y_1(x)$ and $y_2(x)$ are two linear independent solutions of a homogeneous second-order differential equation. Then [16]
	\begin{equation}
	\begin{array}{c}
		\displaystyle\int\limits_{x_i}^x \frac{W\{y_1, y_2\}}{\left(A_1 y_1(t)+A_2 y_2(t)\right)^2}dt = - \frac1{A_1^2+A_2^2} \left(\frac{A_2 y_1(x)- A_1 y_2(x)}{A_1 y_1(x)+A_2 y_2(x)}- \frac{A_2 y_1(x_i)- A_1 y_2(x_i)}{A_1 y_1(x_i)+A_2 y_2(x_i)}\right)
  	\end{array}
	\label{susy-article-eqs:21}
	\end{equation}
where $A_1$, $A_2>0$ and $W\{y_1, y_2\}= y_1 y'_2 - y'_1y_2$ is Wronskian, which is independent of $x$ for the second-order differential equation, reduced to a canonical form. Therefore it can be taken out from under the sign of the integral. This relation is very useful for calculating the integrals included in considering formalism. 

We begin our consideration based on the case $c=1$ in the expression for the general solution of the auxiliary equation (\ref{susy-article-eqs:13}). In expression (\ref{susy-article-eqs:15}) for the superpotential, it is natural to set $x_i=-\infty$, since 
\[\displaystyle \tilde \varphi(x, \varepsilon, c = 1)= \frac{{N^{- 1}}}{{\varphi_1(x, \varepsilon)+\varphi_2(x, \varepsilon)}}\] in this limit it tends to zero. Therefore
	\begin{equation}
		\begin{array}{c}
		\displaystyle N^{- 2}\int\limits_{- \infty}^x{\frac{{dt}}{{\left({\varphi_1(t, \varepsilon)+c\;\varphi_2(t, \varepsilon)}\right)^2}}}=\\
		\displaystyle = - \frac{{N^{- 2}}}{{\;(1+c^2)\;W\{\varphi_1, \varphi_2 \}}}\left[{\Delta(x, \varepsilon, c)- \Delta(- \infty, \varepsilon, c)}\right] \\ 
		\displaystyle N^{- 2}= - \frac{1}{{\;(1+c^2)\;W\{\varphi_1, \varphi_2 \}}}\left[{\Delta(+\infty, \varepsilon, c)- \Delta(- \infty, \varepsilon, c)}\right] \\
		\displaystyle \Delta(x, \varepsilon, c)= \frac{{c\varphi_1(x, \varepsilon)- \varphi_2(x, \varepsilon)}}{{\varphi_1(x, \varepsilon)+c\varphi_2(x, \varepsilon)}}
		\end{array}
	\label{susy-article-eqs:22}
	\end{equation}
Using the above relations for $c=1$, it is easy to obtain an expression for the superpotential up to a constant term.
	\begin{equation}
		\begin{array}{c}
		\displaystyle W(x, \varepsilon, \lambda)= - \frac{1}{2}\ln \left({1+\lambda N^{- 2}\int\limits_{- \infty}^x{\frac{{dt}}{{\left({\varphi_1(x, \varepsilon)+\varphi_2(x, \varepsilon)}\right)^2}}}}\right)=\\
		\\
		\displaystyle = - \frac{1}{2}\ln \left({\frac{{\varphi_1(x, \varepsilon)+\Lambda(\varepsilon, \lambda)\varphi_2(x, \varepsilon)}}{{\varphi_1(x, \varepsilon)+\varphi_2(x, \varepsilon)}}}\right)\\
		\\
		\displaystyle \Lambda(\varepsilon, \lambda)= \frac{{\Delta(\infty, \varepsilon, c = 1)- \lambda -(\lambda+1)\;\Delta(- \infty, \varepsilon, c = 1)}}{{\Delta(\infty, \varepsilon, c = 1)+\lambda -(\lambda+1)\;\Delta(- \infty, \varepsilon, c = 1)}}
		\label{susy-article-eqs:23}
		\end{array}
	\end{equation}
Note that the quantities $\Delta(\pm\infty, \varepsilon, \lambda)$ included in $\Lambda(\varepsilon, \lambda)$ are determined by the asymptotic behavior of the solutions of the auxiliary equation. This shows that if in the case $H_-^-$ the potential is determined by a symmetric combination $\varphi_1(x, \varepsilon)+\varphi_2(x, \varepsilon)$, then in the case $H_+^+$---asymmetric  combination $\varphi_1(x, \varepsilon)+\Lambda(\varepsilon, \lambda)\;\varphi_2(x, \varepsilon)$:
	\begin{equation}
		H_-^-= H_+^- - \frac{{d^2}}{{dx^2}}\ln(\varphi_1(x, \varepsilon)+\varphi_2(x, \varepsilon))
	\label{susy-article-eqs:24}
	\end{equation}
	\begin{equation}
		H_+^+= H_+^- - \frac{{d^2}}{{dx^2}}\ln \left({\varphi_1(x, \varepsilon)+\Lambda(\varepsilon, \lambda)\;\varphi_2(x, \varepsilon)}\right)
	\label{susy-article-eqs:25}
	\end{equation}
In a manner, the potentials $U_-^-(x, \varepsilon)$ and $U_+^+(x, \varepsilon, \lambda)$ are form-invariant [17]. Those. potentials and wave functions are obtained from each other by changing parameters, and the spectra of their Hamiltonians are identical. Moreover, this takes place regardless of the choice of the initial Hamiltonian. As will be shown in the next section, if the potential in $H_-^-$ is a multi-well symmetric potential, then in $H_+^+$ it is asymmetric. Note that a change in the shape of the latter is possible by varying both, $\varepsilon$ and $\lambda$.
\subsection{Isospectral Hamiltonians with almost equidistant spectrum}
In order to obtain specific expressions for the potentials and wave functions, we choose the Hamiltonian with the potential of a harmonic oscillator (HO) as the initial Hamiltonian. Consider \[\displaystyle \varepsilon<E_0 = \frac{\omega}{2}, (\hbar=m=1)\] the solution of the auxiliary equation for 
	\begin{equation}
		\left({\frac{{d^2}}{{dx^2}}+2(\varepsilon - \frac{{\omega^2 x^2}}{2})}\right)\varphi(x, \varepsilon)= 0
	\label{susy-article-eqs:31}
	\end{equation}
Introducing dimensionless variable $\xi = \sqrt{2\omega}x$, we obtain the equation for $\varphi(\xi, \bar\varepsilon)$, 
	\begin{equation}
		\left({\frac{{d^2}}{{d\xi^2}}+(\nu+\frac{1}{2}- \frac{{\xi^2}}{4})}\right)\;\varphi(\xi, \bar \varepsilon)= 0\;, \;\nu = - \frac{1}{2}+\bar \varepsilon
	\label{susy-article-eqs:32}
	\end{equation}
The above equation has two linearly independent solutions: $D_\nu(\sqrt 2 \xi), ~D_\nu(-\sqrt 2 \xi)$. These are solutions that represent the functions of a parabolic cylinder. According to the terminology adopted earlier, let's put $\varphi_1(\xi, \bar\varepsilon)= D_\nu(\sqrt 2 \xi), ~\varphi_2(\xi, \bar\varepsilon)= D_\nu(-\sqrt 2 \xi)$. Wronskian [18] \[\displaystyle W\{\varphi_1, \varphi_2 \}= \frac{{2\sqrt \pi}}{{\Gamma(- \nu)}}\], where $\Gamma(-\nu)$ is the gamma function. The symmetrical solution of the auxiliary equation is chosen in the form:
	\begin{equation}
		\varphi(\xi, \bar \varepsilon, 1)= D_\nu(\sqrt 2 \xi)+D_\nu(- \sqrt 2 \xi)
	\label{susy-article-eqs:33}
	\end{equation}
As can be seen from (\ref{susy-article-eqs:33}) $\varphi(x, \bar \varepsilon, 1)$ is an even function of $\xi$. To obtain the explicit form of the superpotential, it is necessary to calculate the integral included in the definition of $W(x, \varepsilon, \lambda)$ and the normalization constant $N^{-2}$ using (\ref{susy-article-eqs:22}). Due to symmetry of function $\varphi(\xi, \bar\varepsilon,1)$, the expression for the integral is simplified and has the form:
	\begin{equation}
		\displaystyle 1+\lambda N^{- 2}\int\limits_{- \infty}^\xi{\frac{{dt}}{{\left({\varphi_1+\varphi_2}\right)^2}}}= 1 - \frac{{\lambda N^{- 2}}}{{2W\{\varphi_1, \varphi_2 \}}}\left[{\left({\frac{{\varphi_1(\xi, \bar \varepsilon)- \varphi_2(\xi, \bar \varepsilon)}}{{\varphi_1(\xi, \bar \varepsilon)+\varphi_2(\xi, \bar \varepsilon)}}}\right)+\Delta(+\infty, \bar \varepsilon, 1)}\right]
	\label{susy-article-eqs:34}
	\end{equation}
Therefore, the superpotential and the normalization constant can be represented as:
	\begin{align}
		\displaystyle W\{\xi, \bar \varepsilon, \lambda \}&= \ln \left({\frac{{(1+\frac{\lambda}{2}+\frac{\lambda}{{2\Delta(+\infty, \bar \varepsilon, 1)}})\varphi_1+(1+\frac{\lambda}{2}- \frac{\lambda}{{2\Delta(+\infty, \bar \varepsilon, 1)}})\varphi_2}}{{\varphi_1+\varphi_2}}}\right)
	\label{susy-article-eqs:35}\\
	\nonumber
		\displaystyle N^{- 2}&= - \frac{{W\{\varphi_1, \varphi_2 \}}}{{\Delta(+\infty, \bar \varepsilon, 1)}}= \frac{{2\sqrt \pi}}{{\Gamma(- \nu)}}
	\end{align}
Using the asymptotic value of the functions of a parabolic cylinder, we obtain $\Delta(+\infty, \bar\varepsilon, 1)=-1$. Then the expressions for the super potential take on a rather compact form:
	\begin{equation}
		\displaystyle W(\xi, \bar \varepsilon, \lambda)= - \frac{1}{2}\ln \left({\frac{{\varphi_1(\xi, \bar \varepsilon)+(1+\lambda)\varphi_2(\xi, \bar \varepsilon)}}{{\varphi_1(\xi, \bar \varepsilon)+\varphi_2(\xi, \bar \varepsilon)}}}\right)= - \frac{1}{2}\ln \left({\frac{{\varphi(\xi, \bar \varepsilon, 1+\lambda)}}{{\varphi(\xi, \bar \varepsilon, 1)}}}\right)
	\label{susy-article-eqs:36}
	\end{equation}
From (\ref{susy-article-eqs:36})) it is seen that the parameter value of $\lambda$ is limited by the condition $\lambda>-1$. Substituting into the expressions obtained in Section \ref{5_2}, the value of the superpotential (\ref{susy-article-eqs:36}) we obtain expressions for $H_-^-$ and $\psi_-^-(x, E)$ - the wave functions of HO:
	\begin{equation}
		\begin{array}{c}
		\displaystyle H_-^- = H_+^- - \frac{{d^2}}{{dx^2}}\ln \left({D_\nu(\sqrt 2 \xi)+D_\nu(- \sqrt 2 \xi)}\right),\\
		\displaystyle \psi_-^-(\xi, E_i)= \frac{1}{{\sqrt{2(E_i - \bar \varepsilon)}}}\frac{{W\left\{{\psi_+^-(\xi, E_i), \varphi(\xi, \bar \varepsilon, 1)}\right\}}}{{\varphi(\xi, \bar \varepsilon, 1)}}\\
		\displaystyle \psi_-^-(x, E = 0)= \frac{{N^{- 1}}}{{\left({D_\nu(\sqrt 2 \xi)+D_\nu(- \sqrt 2 \xi)}\right)}}= \frac{{N^{- 1}}}{{\varphi(x, \bar \varepsilon, 1)}}, \quad \;N^{- 2}= \frac{{2\sqrt \pi}}{{\Gamma(- \nu)}}
		\end{array}
	\label{susy-article-eqs:37}
	\end{equation}
Similarly, we consider the relationship of $H_+^+$ and $\psi_+^+(\xi, E_n)$ with $H_+^-$ and $\psi_+^-(\xi, E_n)$,  respectively. Using (\ref{susy-article-eqs:36}) we obtain the following expressions for $H_+^+$ and $\psi_+^+(\xi, E_n)$:
	\begin{equation}
		\begin{array}{c}
		\displaystyle H_+^+= H_+^- - \frac{{d^2}}{{dx^2}}\ln \left({D_\nu(\sqrt 2 \xi)+(1+\lambda)D_\nu(- \sqrt 2 \xi)}\right),\\
		\\
		\displaystyle \psi_+^+(\xi, E_i)= \frac{1}{{\sqrt{2(E_i - \bar \varepsilon)}}}\frac{{W\left\{{\psi_+^-(\xi, E_i), \varphi(\xi, \bar \varepsilon, \lambda+1)}\right\}}}{{\varphi(\xi, \bar \varepsilon, \lambda+1)}}\\
		\\
		\displaystyle \psi_+^+(\xi, E = 0)= \frac{{N_{\lambda+1}^{- 1}}}{{\left({D_\nu(\sqrt 2 \xi)+(\lambda+1)D_\nu(- \sqrt 2 \xi)}\right)}}= \frac{{N_{\lambda+1}^{- 1}}}{{\varphi(x, \bar \varepsilon, \lambda+1)}},\\
		\\
		\displaystyle N_{\lambda+1}^{- 2}= \frac{{2(\lambda+1)\sqrt \pi}}{{\Gamma(- \nu)}}
		\end{array}
	\label{susy-article-eqs:38}
	\end{equation}
Note that the only way of varying the shape of the potential in terms of dimensionless variables $\xi$ is to vary parameters $\Lambda$ and $\bar\varepsilon$ in the range $\bar\varepsilon<1/2,\ \Lambda>0$. In the natural units $x$, there is one more parameter $\omega$ for variation.  
\begin{figure}
\begin{center}
 \includegraphics[width=0.5\linewidth]{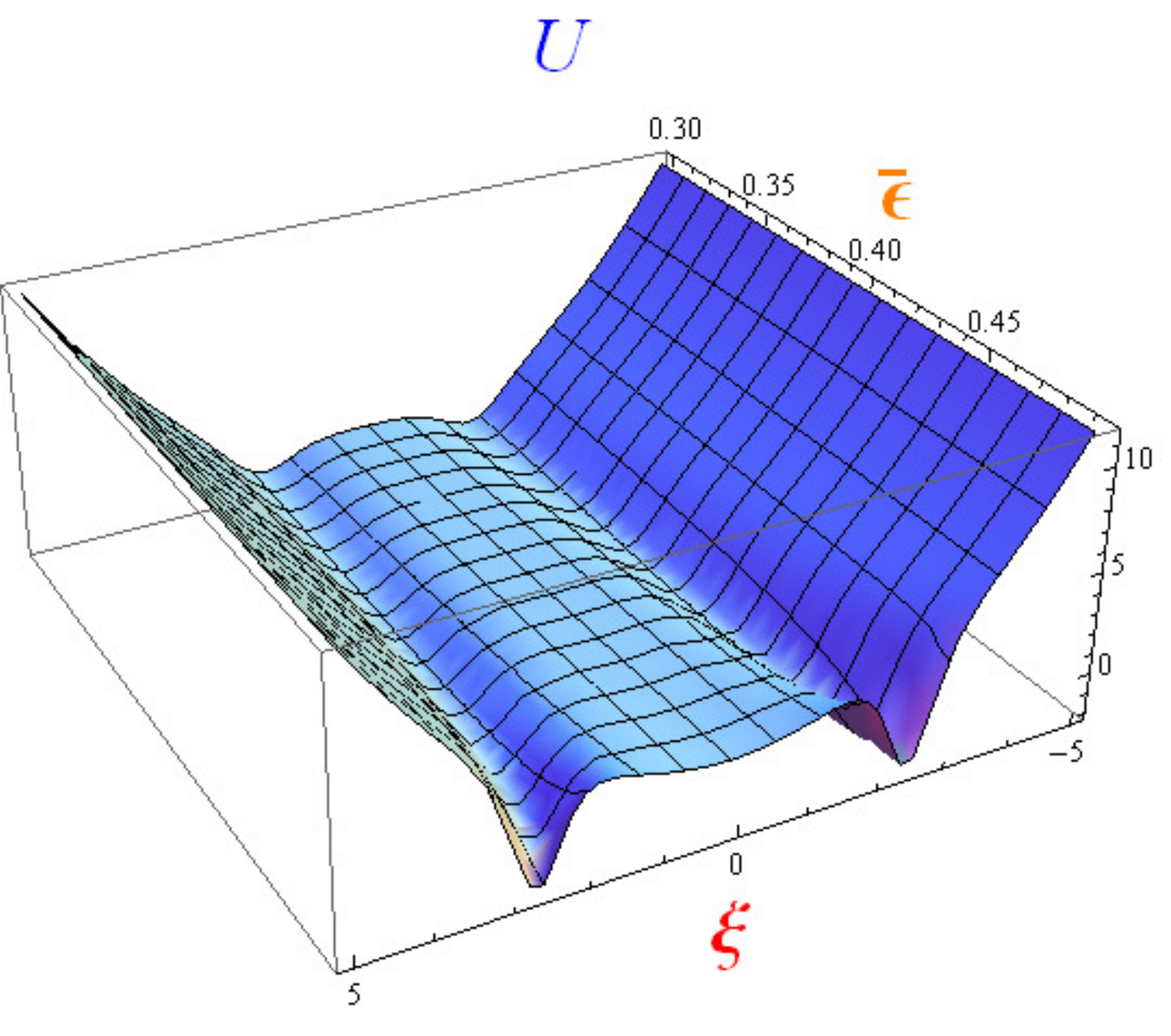}
  \caption{$\bar\varepsilon$-dependence of potential $U_-^-(\xi, \bar \varepsilon, \lambda = 0)$.\label{susy-article-fig:1}}
\end{center}
\end{figure}

\begin{figure}
\begin{center}
 \includegraphics[width=0.5\linewidth]{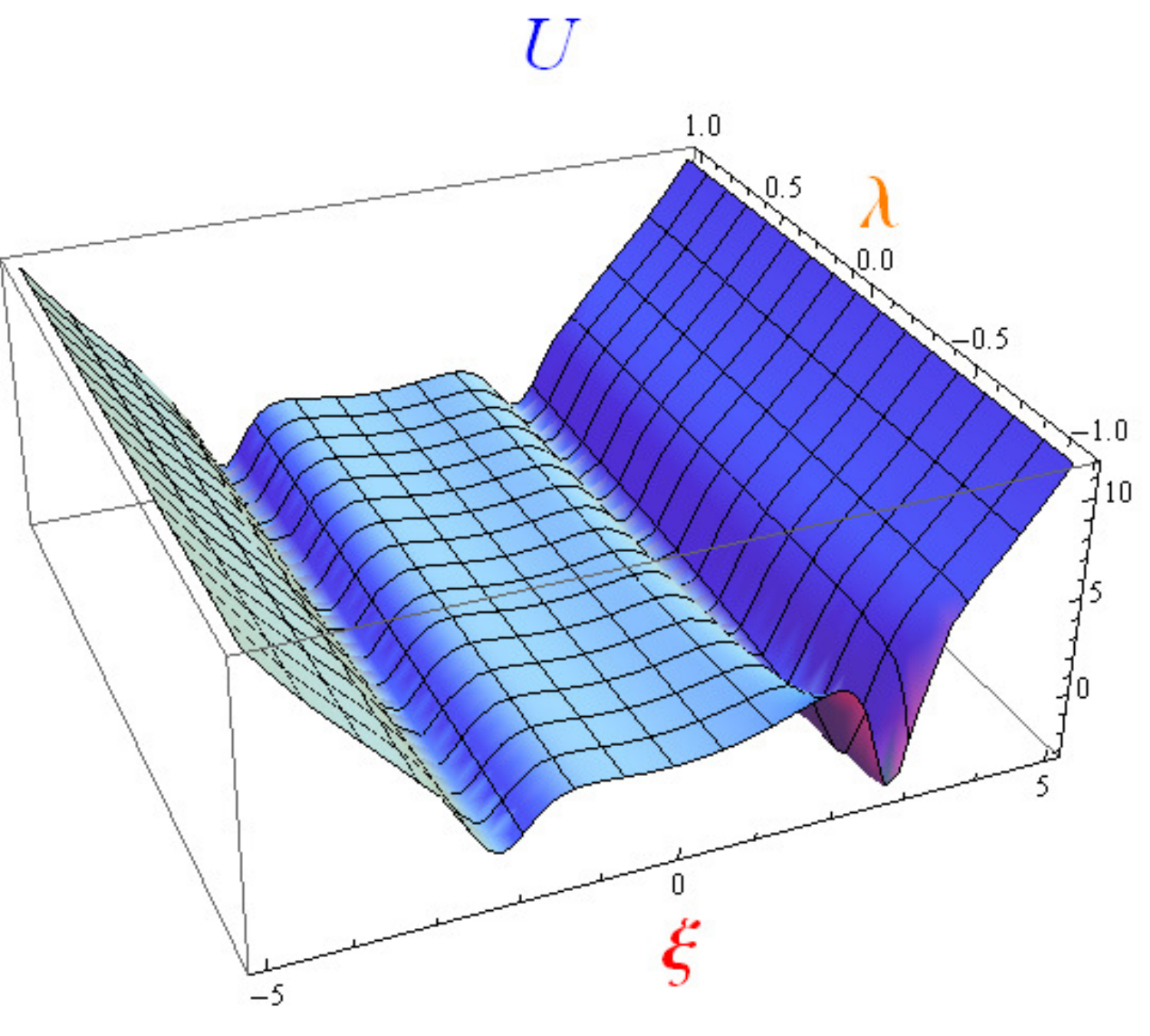}
  \caption{$\lambda$-dependence of potential $U_+^+(\xi, \bar \varepsilon = 0.47, \lambda)$.\label{susy-article-fig:2}}
\end{center}
\end{figure}
So, in this section using the example of the harmonic oscillator  Hamiltonian, isospectral Hamiltonians with multi-well potentials and the corresponding wave functions are obtained in explicit form. The presence of parameters allows you to change the shape of the potentials in a wide range (see Fig.\ref{susy-article-fig:1}. and Fig.\ref{susy-article-fig:2}). This is especially important when studying such phenomena as tunneling [19], which are sensitive to structural features of multi-well potentials, and also when using wave functions (\ref{susy-article-eqs:37}) and (\ref{susy-article-eqs:38}) as a basis for diagonalization of Hamiltonians with phenomenological potentials.
\subsection{Construction of isospectral Hamiltonians with triple-well potentials.}
The Hamiltonians obtained with double-well potentials obtained above have a significant disadvantages. The condition for the appearance of several local minima in the potential $(E_0-\varepsilon) \ll E_0$ impress limitations on the choice of parameters of corresponding minima. This, in particular, does not allow us to consider cases of arbitrary arrangement of the tunneling doublet relative to the ground state of the initial Hamiltonian. To eliminate this limitation, it is necessary to repeat the above adding procedure an additional level below the ground state of Hamiltonians (\ref{susy-article-eqs:37}) and (\ref{susy-article-eqs:38}).

The procedure for constructing exactly solvable models from initial Hamiltonian $H_0$ associated with adding levels below the ground state is proposed in [20,21] (Crum-Krein method) and it was repeatedly discussed in later works (see, for example, [22,23]). To construct Hamiltonians with triple-well potentials, we will use the results to obtain exactly solvable models with two-well potentials in extended supersymmetric quantum mechanics (SQM) [32,33]. It should be noted that such a consideration is also possible within the framework of the Crum-Krein method, although some points (for example, normalization of the wave functions of additional states) require a separate consideration in this approach. Therefore, when constructing isospectral Hamiltonians with multi-well potentials, we prefer the method of successively applying supersymmetry transformations to solutions of the original Hamiltonian $H_0$.

Relations (\ref{susy-article-eqs:17}), (\ref{susy-article-eqs:17}), (\ref{susy-article-eqs:24}) and (\ref{susy-article-eqs:25}) are the initial ones for the further construction of isospectral Hamiltonians with triple-well potentials. We consider as a new initial Hamiltonian $H^+_+$ with a two-well potential
\begin{equation}
    \label{523}
{\tilde H}_0\equiv H^+_+=H^-_+-\frac{d^2}{dx^2}\ln \big(\varphi_1(x,\varepsilon)+\Lambda(\varepsilon,\lambda)\varphi_2(x,\varepsilon)\big).
\end{equation}
Recall that the spectrum ${\tilde H}_0$ contains the states of the initial Hamiltonian $H_0$ and an additional level with energy $\varepsilon<E_0$, i.e. is the energy of the ground state of Hamiltonian ${\tilde H}_0$. We consider the solutions of the equation ${\tilde H}_0\chi(x)=\varepsilon_1\chi(x)$ for $\varepsilon_1<\varepsilon$. There are two linearly independent solutions (non-normalization):
\begin{equation}
    \label{524}
    \chi_1(x,\varepsilon_1,\varepsilon,\Lambda)=\frac{W[\varphi_1(x,\varepsilon_1),\varphi(x,\varepsilon,\Lambda)}{\varphi(x,\varepsilon,\Lambda)};\quad 
    \chi_2(x,\varepsilon_1,\varepsilon,\Lambda)=-\frac{W[\varphi_2(x,\varepsilon_1),\varphi(x,\varepsilon,\Lambda)}{\varphi(x,\varepsilon,\Lambda)}
\end{equation}
where $\varphi_i(x,\varepsilon_1)$, $i=1,2$ are the solutions of the equation $H_0\varphi_i(x,\varepsilon_1)=\varepsilon_1\varphi_i(x,\varepsilon_1)$ and  $\varphi(x,\varepsilon,\Lambda)=\varphi_1(x,\varepsilon_1)+\Lambda(\varepsilon,\lambda)\varphi_2(x,\varepsilon_1)$. Note that $\chi_i(x,\varepsilon_1,\varepsilon,\Lambda)$, $i=1,2$ are linearly independent solutions that are non-negative and have asymptotic behavior for $x\to-\infty\Rightarrow\chi_1(x)\to+\infty\ (\chi_2(x)\to0$, and for $x\to+\infty\Rightarrow\chi_1(x)\to0\ (\chi_2(x)\to+\infty$. The presence of the minus sign in the definition of $\chi_2(x,\varepsilon_1)$ ensures the correct asymptotic behavior of non-normalized functions. With a certain choice of constants, the general solution $\chi(x,\varepsilon_1,\varepsilon,\Lambda,\tilde c)=\tilde N(\chi_1(x,\varepsilon_1,\varepsilon,\Lambda)+\tilde c \chi_2(x,\varepsilon_1,\varepsilon,\Lambda))$ (the common constant that will be used for normalization) does not have nodes on the entire axis. In further consideration, we will assume $\tilde c=1$. Then the function \[\tilde\chi(x,\varepsilon_1,\varepsilon,\Lambda,\tilde c)=\frac{\tilde N^{-1}}{\chi_1(x,\varepsilon_1,\varepsilon,\Lambda)+\tilde c \chi_2(x,\varepsilon_1,\varepsilon,\Lambda)}\] is finite and can be normalized to unity for each specific value of the parameters.
Assuming
\begin{align}
    \label{525}
    \tilde H_0-\varepsilon_1\equiv H_+^+&=\tilde H_+^-(x,p)=\frac12 p^2 + \frac12 \left[(W'(x,\varepsilon_1,\varepsilon,\Lambda))^2+W''(x,\varepsilon_1,\varepsilon,\Lambda)\right],\\
    \nonumber
    W(x,\varepsilon_1,\varepsilon,\Lambda)&=\ln\chi(x,\varepsilon_1,\varepsilon,\Lambda,1) = \ln\frac{W[\phi(x,\varepsilon_1,1),\varphi(x,\varepsilon,\Lambda)}{\varphi(x,\varepsilon,\Lambda)}
\end{align}where $\phi(x,\varepsilon_1,1)=\varphi_1(x,\varepsilon_1)-\varphi_2(x,\varepsilon_1)$. The spectrum of the Hamiltonian $\tilde H_+^-$ completely coincides with the spectrum of the Hamiltonian $H_+^+$ with a double-well potential, in which, in addition to states of $H_0$, there is an additional level with energy $\varepsilon<E_0$. Superpartner of the Hamiltonian $\tilde H_+^-$ is the Hamiltonian $\tilde H_-^-$ with the following form
\begin{equation}
    \label{526}
    \tilde H_-^- = \tilde H_+^- - \frac{d^2}{dx^2}\ln\chi(x,\varepsilon_1,\varepsilon,\Lambda,1)= H_0 -\frac{d^2}{dx^2} \ln W[\phi(x,\varepsilon_1,1),\varphi(x,\varepsilon,\Lambda)]
\end{equation}
The spectrum of Hamiltonian (\ref{526}) consists of states $H_0$ and additional levels with energies $\varepsilon$ and $\varepsilon_1$. The corresponding wave functions have the form
\begin{align}
    \label{527a}
\Psi_0(x,\varepsilon_1;\varepsilon,\Lambda)&=\frac{\tilde N^{-1}}{\chi_1(x,\varepsilon_1,\varepsilon,\Lambda)+\tilde c \chi_2(x,\varepsilon_1,\varepsilon,\Lambda)} = \frac{\tilde N^{-1}\varphi(x,\varepsilon,\Lambda)}{W[\phi(x,\varepsilon_1,1), \varphi(x,\varepsilon,\Lambda)]}\\
    \label{527b}
\Psi_1(x,\varepsilon;\varepsilon_1,\Lambda)&=\frac{N^{-1}}{\sqrt{2(\varepsilon-\varepsilon_1)}}\left(\frac{d}{dx}-\frac{\chi'(x,\varepsilon_1,\varepsilon,\Lambda,1)}{\chi(x,\varepsilon_1,\varepsilon,\Lambda,1)}\right)\tilde\varphi(x,\varepsilon,\Lambda) =\\
\nonumber &=\sqrt{2(\varepsilon-\varepsilon_1)}\frac{N^{-1}\phi(x,\varepsilon_1,1)}{W[\phi(x,\varepsilon_1,1), \varphi(x,\varepsilon,\Lambda)]}\\
    \label{527c}
\Psi_-^-(x,E_i)&=\frac{1}{\sqrt{2(E_i-\varepsilon_1)}}\left(\frac{d}{dx}-\frac{\chi'(x,\varepsilon_1,\varepsilon,\Lambda,1)}{\chi(x,\varepsilon_1,\varepsilon,\Lambda,1)}\right)\psi_+^+(x,E_i) =\\
\nonumber
&=\sqrt{\frac{E_i-\varepsilon}{E_i-\varepsilon_1}}\left(\psi_+^-(x,E_i)+\frac{\varepsilon-\varepsilon_1}{E_i-\varepsilon_1}\phi(x,\varepsilon_1)\frac{W[\psi_+^-(x,E_i),\varphi(x,\varepsilon,\Lambda)]}{W[\phi(x,\varepsilon,1),\varphi(x,\varepsilon,\Lambda)]}\right)
\end{align}
Thus, the spectrum of the Hamiltonian $\tilde H_-^-$ consists of the states of the initial Hamiltonian $H_0$ and two additional levels with energies $\varepsilon$ and $\varepsilon_1$ ($\varepsilon_1<\varepsilon<E_0$), the wave functions of which are given by relations (\ref{527a},\ref{527b},\ref{527c}). It is easy to calculate the normalization constant of the wave function of the ground state $\Psi_0(x,\varepsilon_1,\varepsilon,\Lambda)$ using (\ref{susy-article-eqs:22}) with allowance for the asymptotic behavior of the functions $\chi_i(x,\varepsilon_1,\varepsilon,\Lambda)$, $i=1,2$. Then:
\begin{equation}
    \label{528}
\tilde N^{-2}=W[\chi_1,\chi_2]\equiv2(\varepsilon-\varepsilon_1)W[\varphi_1(x,\varepsilon_1),\varphi_1(x,\varepsilon_1)]    
\end{equation}
For a more symmetric representation of the wave functions of the added levels, it is natural to redefine the normalization constant $\tilde N^{-2}\to2(\varepsilon-\varepsilon_1)N^{=2}=2(\varepsilon-\varepsilon_1)W[\varphi_1(x,\varepsilon_1),\varphi_1(x,\varepsilon_1)]$. Then the expression (\ref{527b}) with a modified normalization constant will have the following form:  
\[
\Psi_1(x,\varepsilon;\varepsilon_1,\Lambda)=\frac{N^{-1}\phi(x,\varepsilon_1,1)}{W[\phi(x,\varepsilon_1,1), \varphi(x,\varepsilon,\Lambda)]}
\]
Similarly to the previously considered case of isospectral Hamiltonians with double-well potentials, we establish the form-invariance property for the potentials in $\tilde H_-^-$ and $\tilde H_+^+$. The connection between the Hamiltonians $\tilde H_+^+$ and $\tilde H_+^-$ reads:
\begin{equation}
    \label{529}
    \tilde H_+^+=\tilde H_+^- -\frac{d^2}{dx^2}\ln\left(\frac{\tilde\chi(x,\varepsilon_1,\varepsilon,\Lambda,1)}{1+\tilde\lambda\int\limits^x_{x_i}dt\tilde\chi^2(t,\varepsilon_1,\varepsilon,\Lambda,1)}\right)
\end{equation}
Using relation (\ref{susy-article-eqs:22}) it is easy to show the validity of the following equality:
\begin{equation}
    \label{530}
    \frac{\tilde\chi(x,\varepsilon_1,\varepsilon,\Lambda,1)}{1+\tilde\lambda\int\limits^x_{x_i}dt\tilde\chi^2(t,\varepsilon_1,\varepsilon,\Lambda,1)} = \frac{\tilde N^{-1}_{\Lambda_1}}{\chi_1(x,\varepsilon_1,\varepsilon,\Lambda)+\Lambda_1\chi_2(x,\varepsilon_1,\varepsilon,\Lambda)},
\end{equation}
where $\tilde N^{-2}_{\Lambda_1} = (\tilde\lambda+1)\tilde N^{-2}$, $\Lambda_1=(\tilde\lambda+1)$, $\tilde\lambda>-1$. Consequently,
\begin{equation}
    \label{531}
    \tilde H_+^+=\tilde H_+^- -\frac{d^2}{dx^2}\ln\left|\chi(x,\varepsilon_1,\varepsilon,\Lambda_1,\Lambda)\right|=H_0-\frac{d^2}{dx^2}\ln W[\phi(x,\varepsilon_1,\Lambda_1),\varphi(x,\varepsilon,\Lambda)].    
\end{equation}
The wave functions corresponding to the states of $\tilde H^+_+$ are obtained from relations (\ref{527a},\ref{527b},\ref{527c}) by replacement $\phi(x,\varepsilon_1,1)\to\phi(x,\varepsilon_1,\Lambda_1)=\phi_1(x,\varepsilon_1)-\Lambda_1\varphi_2(x,\varepsilon_1)$:
\begin{align}
    \label{532a}
\Psi_0(x,\varepsilon_1,\Lambda_1;\varepsilon,\Lambda)&= \frac{\tilde N^{-1}_{\Lambda_1}\varphi(x,\varepsilon,\Lambda)}{W[\phi(x,\varepsilon_1,\Lambda_1), \varphi(x,\varepsilon,\Lambda)]}\\
    \label{532b}
\Psi_0(x,\varepsilon,\Lambda;\varepsilon,\Lambda_1)&= \frac{ N^{-1}\varphi(x,\varepsilon_1,\Lambda_1)}{W[\phi(x,\varepsilon_1,\Lambda_1), \varphi(x,\varepsilon,\Lambda)]}\\
    \label{532c}
\Psi_+^+(x,E_i)&= \sqrt{\frac{E_i-\varepsilon}{E_i-\varepsilon_1}}\left(\psi_+^-(x,E_i)+\frac{\varepsilon-\varepsilon_1}{E_i-\varepsilon_1}\phi(x,\varepsilon_1,\Lambda_1)\frac{W[\psi_+^-(x,E_i),\varphi(x,\varepsilon,\Lambda)]}{W[\phi(x,\varepsilon_1,\Lambda_1),\varphi(x,\varepsilon,\Lambda)]}\right)
\end{align}
It should be noted that the relations obtained above include only the wave functions of the Hamiltonian $H_0$, as well as non-normalizable solutions of the Schrödinger equation $H_0\varphi(x)=\varepsilon\varphi(x)$ at energies below the ground state energy of $H_0$. The potentials that are part of the Hamiltonians $\tilde H_-^-$ and $\tilde H_+^+$ contain a number of parameters ($\varepsilon\varepsilon_1,\Lambda,\Lambda_1$), the change of which allows you to modify their shape in a wide range from symmetric to substantially deformed. Obtained relations are the basis [24] of constructing a basis for the study of Hamiltonians with multi-well (two and three well) potentials.
\subsection{Realization of the basis on the  harmonic oscillator foundation.}
In the future, we will use an exactly solvable model with a multiwell potential constructed on the basis of the HO model. In this case, the unnormalized solutions for $\varepsilon_j<E_0$ are given in terms of the parabolic cylinder functions
\[
\varphi_1(\xi, \bar\varepsilon) = D_\nu(\sqrt2\xi),\ \varphi_2(\xi, \bar\varepsilon) = D_\nu(-\sqrt2\xi), \quad \xi=\sqrt\omega x,\ \nu=-\frac12+\bar\varepsilon,\ \bar\varepsilon=\frac{\varepsilon}{\omega},
\]
\[
\varphi_1(\xi, \bar\varepsilon_1) = D_\mu(\sqrt2\xi),\ \varphi_2(\xi, \bar\varepsilon_1) = D_\mu(-\sqrt2\xi),  \quad \nu=-\frac12+\bar\varepsilon_1,\ \bar\varepsilon_1=\frac{\varepsilon_1}{\omega},
\]
Normalization constants of wave functions of the added states are:
\[
N^{-2}=4(\nu-\mu)\frac{\sqrt{\pi}}{\Gamma(-\nu)},\tilde N^{-2}_{\Lambda_1}=\Lambda_1\tilde N^{-2}, \quad \Lambda_1>0. 
\]
\[
\tilde N^{-2}=4(\nu-\mu)\frac{\sqrt{\pi}}{\Gamma(-\mu)}, N^{-2}_\Lambda=\Lambda N^{-2}, \quad \Lambda>0, 
\]
Explicit expressions for the wave functions and Hamiltonians are obtained from relations  (\ref{531}) and (\ref{532a},\ref{532b},\ref{532c}) by substituting $\varphi(x,\varepsilon,\Lambda)\to D_\nu(\sqrt2\xi) +\Lambda D_\nu(-\sqrt2\xi)$, $\phi(x,\varepsilon_1,\Lambda_1)\to D_\mu(\sqrt2\xi) -\Lambda_1 D_\mu(-\sqrt2\xi)$, and $\psi^-_+(x,E_i)$ are the eigenfunctions of the HO Hamiltonian $H(p,x)=\omega H(p_\xi,\xi)$, where $p_\xi$ and $\xi$ are the dimensionless operators of momentum and coordinates, respectively. So we get the Hamiltonian families. It should be noted that in terms of the dimensionless variable $\xi$ the way to vary the form of the potential is by varying $\bar\varepsilon$, $\bar\varepsilon_1<1/2$ and $0<\Lambda,\Lambda_1$. In the case of natural units $x$, additionally, the form of the potential (in particular, position of local minima) can be changed by variation $\omega$.

\begin{figure}
    \centering
    \includegraphics[width=0.3\textwidth]{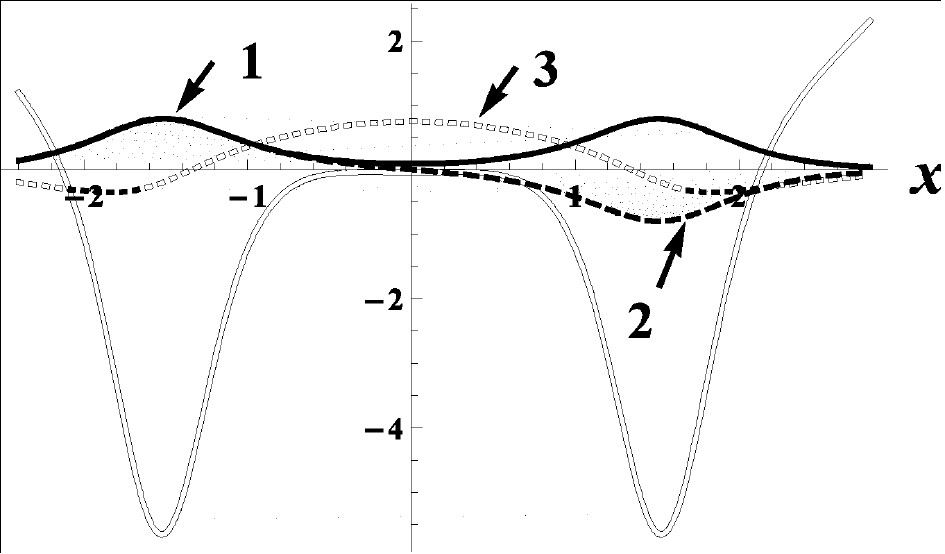}
    \includegraphics[width=0.3\textwidth]{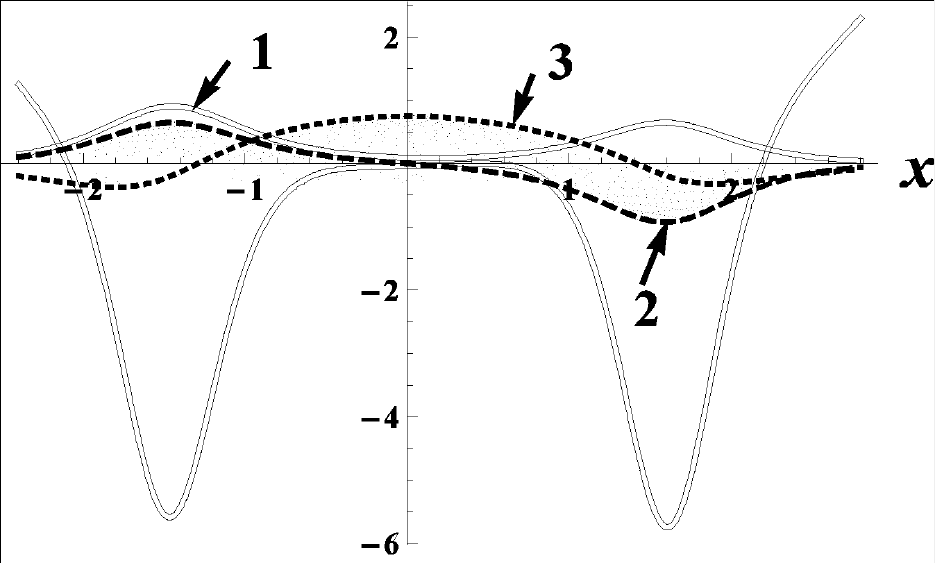}
    \includegraphics[width=0.3\textwidth]{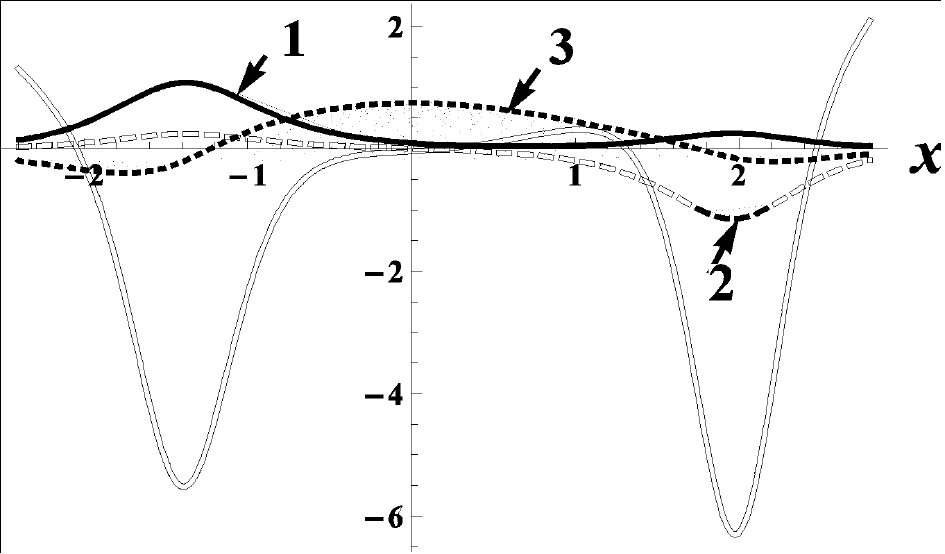}\\
a)\hspace{0.3\textwidth} b)\hspace{0.3\textwidth}  c) 
    \caption{The potential $\tilde U(x,\varepsilon_1,\varepsilon,\Lambda)$ (a--symmetric, b--asymmetric $\Lambda=0.5$, c--asymmetric $\Lambda=0.05$) for $\nu=-3.0$,  $\mu=-3.02$ and lower wave functions (1---ground state, 2,3---first and and second excited states)}
    \label{f3}
\end{figure}

\begin{figure}
    \centering
    \includegraphics[width=0.4\textwidth]{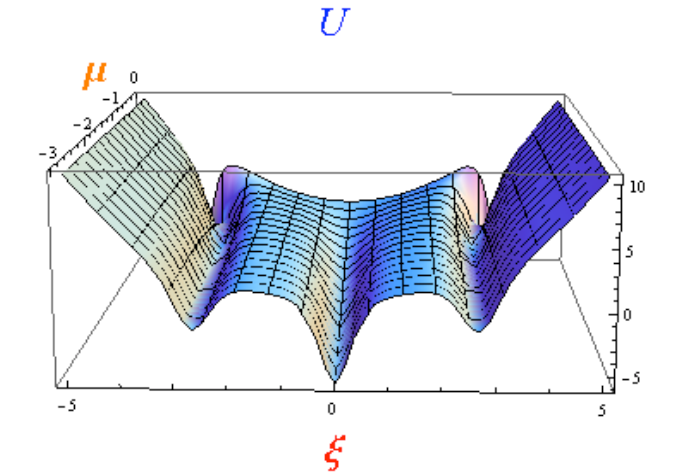}
    \includegraphics[width=0.5\textwidth]{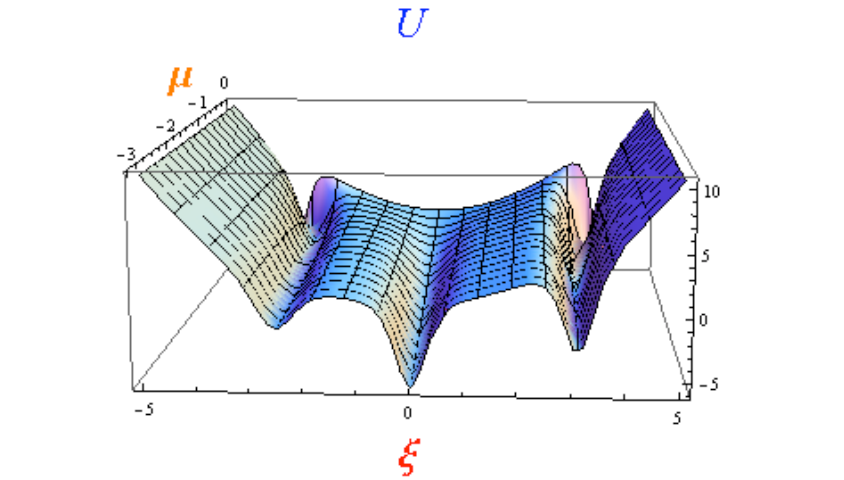}\\
a)\hspace{0.5\textwidth}  b) 
    \caption{Potential $\tilde U^+_+(\xi,\mu,\nu,\Lambda)$ a) $\Lambda=1$, $\nu=-0.02$. b)$\Lambda=0.05$}
    \label{f4}
\end{figure}

\begin{figure}
    \centering
    \includegraphics[width=0.4\textwidth]{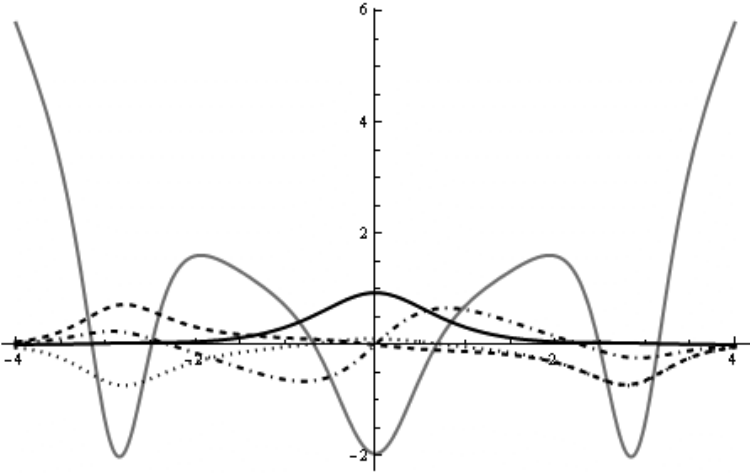}
    \includegraphics[width=0.4\textwidth]{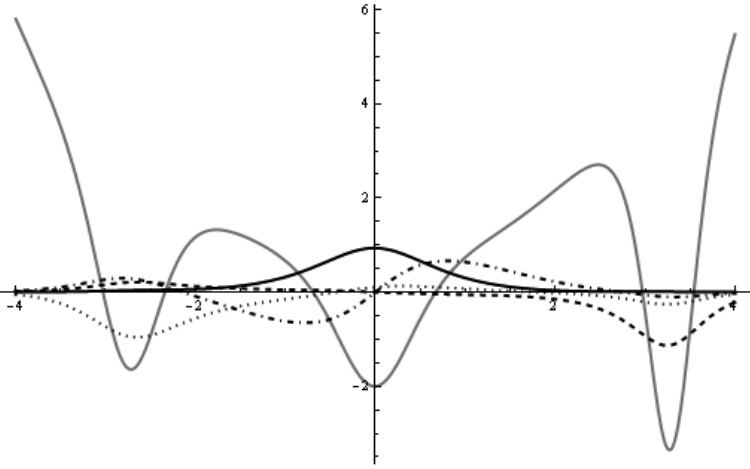}\\
a)\hspace{0.5\textwidth}  b) 
    \caption{Potential $\tilde U^+_+(x,\varepsilon_1,\varepsilon,\Lambda)$ and wave functions. a) $\Lambda=1$, $\mu=-1$, $\nu=-0.02$, b)$\Lambda=0.05$, $\mu=-1$, $\nu=-0.02$.}
    \label{f5}
\end{figure}

A remarkable property of the proposed families of Hamiltonians (\ref{531}) is the fact that they contain both two well and three well potentials. In the case when the tunneling doublet   $\Delta=(\varepsilon-\varepsilon_1)/\omega$ is located at a significant distance from the ground state of the initial Hamiltonian $\varepsilon_0/\omega=1/2$, the potential is two-well (Fig.\ref{f3}). The choice of parameters determining the shape of the reduced potentials guarantees the presence of a larger barrier between local minima, which is important for fulfilling the condition of two-level approximation in the study of tunnel dynamics. In the case of a symmetrical potential (Fig.\ref{f3}a), the behaviour of the central part is similar to the quartic double well potentials often used in the works. By choosing the magnitude of the tunneling doublet and its position, we can approximate the central part of the phenomenological potentials with fairly good accuracy. The asymptotic behaviour of the phenomenological and exactly solvable potentials is different, which leads to a change in the tunnel doublet and sub-barrier wave functions for phenomenological models. So the solutions given in (\ref{532a},\ref{532b},\ref{532c}) can serve as an adequate basis for studying the properties of models with a phenomenological interaction only for the lower spectral states that determine tunnel dynamics. As a rule, the used phenomenological multi-well potentials have asymptotic behaviour $x^n$ ($n$-number of local minima), which leads to significant computational difficulties. The proposed potentials preserve the asymptotic behaviour of the initial potential, independent of the number of local minima.
In the case when the first and second excited states of $\tilde H_+^+$ form a tunnel doublet, and the level of the ground state is removed from them, there is a three-well potential. Figure \ref{f4} shows the dependence of the potential shape upon a change in the energy of the ground state for the symmetric case (a) and with a non-zero deformation parameter of the outside local minima (b,c). In the case of a symmetric potential, as the value  increases, the width of the central local minimum increases, and the outside decreases. In this case, the distance between the outside minima increases. Those. energy is a parameter to control the shape of the potential. In this case, the energies of the first and second excited states remain unchanged. In the presence of deformation  with a decrease , the height of the barrier increases before a deeper minimum.
It is interesting to trace the behavior of the wave functions corresponding to the lower states of the obtained Hamiltonians $\tilde H^-_-$ (sub-barrier states). So for the case when all the wells have the same depth (Fig. \ref{f5}), the fact that the wave function of the ground state (solid curve) is almost completely concentrated in the central part and completely absent in the side wells attracts attention. While the wave functions of the first (dashed) and second (point) excited states practically make small  contribution in the central well.
\section{Conclusion}
We considered possibilities of coherent control of tunneling processes in systems with disturbed reflective symmetry by periodic time-dependent influences. Coherent destruction of tunneling was achieved with the characteristics of the pump field somewhat different from the parameters obtained from the CTD criterion in the two-level approximation for the symmetric case. The number of basic Hamiltonian states, used in the calculations is significantly lower than those using states with a single well potential. This indicates the adequacy of the proposed basis for calculations with phenomenological potentials. Varying the values of the model parameters approximate quite well the form of phenomenological potentials in the sub-barrier region, where the sub-barrier states determine the tunnel dynamics.

In the case of time-dependent deformation parameters, we obtained the effective Hamiltonian, which characterizes the system's response to changes in the shape of the multi-well potential. The time dependence of the Hamiltonian was a bi-chromatic field, the intensity of each component is determined by the degree of change in the probability density of the corresponding orbitals. The presence of such external influences can lead to the effects associated with quantum ratchet. In the future, the question of the possibility of localization of barrier states in systems characterized by the proposed Hamiltonian will be investigated.

\end{document}